\documentclass[epj]{webofc}
\usepackage[varg]{txfonts}   
\usepackage{graphicx,color} 
\usepackage{bm}       
\usepackage{amsmath}  
\usepackage{amssymb}
\usepackage[utf8]{inputenc}
\woctitle{21st International Conference on Few-Body Problems in Physics}
\newcommand{\etal}{\textit{et al.\xspace}}

\begin{document}
\title{Pion's valence-quark GPD and its extension beyond DGLAP region}
\author{ C\'edric Mezrag\inst{1}\fnsep\thanks{\email{cmezrag@anl.gov}} \and 
		Herv\'e Moutarde\inst{2}\fnsep\thanks{\email{herve.moutarde@cea.fr}} \and  
        Jos\'e Rodr\'{\i}guez-Quintero\inst{3}\fnsep\thanks{\email{jose.rodriguez@dfaie.uhu.es}} 
}
\institute{Physics Division, Argonne National Laboratory, Argonne IL 60439,USA             
\and
Centre de Saclay, IRFU/Service de Physique Nucl\'eaire, F-91191 Gif-sur-Yvette, France
\and
Departamento de F\'isica Aplicada, Facultad de CCEE, Universidad de Huelva, Huelva E-21071, Spain           
          }

\abstract{
  We briefly report on a recent computation, with the help of a fruitful algebraic model, sketching the pion valence dressed-quark generalized parton distribution and, very preliminary, discuss on a possible avenue to get reliable results in both Dokshitzer-Gribov-Lipatov-Altarelli-Parisi (DGLAP) and Efremov-Radyushkin-Brodsky-Lepage (ERBL) kinematial regions.
}
\maketitle
\section{Introduction}
\label{intro}

Generalized Parton Distributions (GPDs) were introduced independently by M\"uller \etal \cite{Mueller:1998fv}, Ji \cite{Ji:1996nm} and Radyushkin \cite{Radyushkin:1997ki}. They are related to hadron form factors by sum rules, and contain the usual Parton Distribution Functions (PDFs) as a limiting case. But they not only generalize the classical objects describing the static or dynamical content of hadrons; they also provide unique information about the structure of hadrons, including 3D imaging of its partonic components and access to the quark orbital angular momentum. GPDs have been the object of an intense theoretical and experimental activity ever since (see {\it e.g.} \cite{Tiburzi:2002tq,Theussl:2002xp,Broniowski:2003rp,Ji:2006ea,Broniowski:2007si,Frederico:2009fk} or the reviews \cite{Ji:1998pc,Goeke:2001tz,Diehl:2003ny,Belitsky:2005qn,Boffi:2007yc,Guidal:2013rya} and references therein).

Most of the constraints that apply to GPDs are fulfilled when the GPD is written as a double distribution (DD)
\cite{Mueller:1998fv,Radyushkin:1998es,Radyushkin:1998bz}, which is equivalent to expressing the GPD as a Radon transform \cite{Teryaev:2001qm}. In order to obtain insights into the nature of hadron GPDs, it has been common to model the Radon amplitudes, $F$, $G$, following Refs.\,\cite{Musatov:1999xp}.  This approach has achieved some phenomenological success (see {\it e.g.} \cite{Guidal:2013rya,Mezrag:2013mya}); but more flexible parametrisations enable a better fit to data \cite{Kumericki:2008di}.  Such fits play a valuable role in establishing the GPD framework;
However there is no known parameterization of GPDs relying on first principles only. Computing GPDs in a symmetry-preserving framework is a key ingredient for the {\it a priori} fulfillment of all GPD theoretical constraints.  This observation is highlighted by experience drawn from the simpler case of the pion's valence-quark PDF \cite{Chang:2014lva}.  
In \cite{Mezrag:2014tva,Mezrag:2014jka}, steps are made by following a different approach for the computation of hadron GPDs based on the example provided by the pion's valence-quark PDF. As sketched in \cite{Mezrag:2014jka}, such a procedure only leaves with a reliable result near the so-called forward limit ($\xi=0$), within the DGLAP region~\cite{Dokshitzer:1977sg,Gribov:1972ri,Lipatov:1974qm,Altarelli:1977zs}. We will now preliminary discuss how, mainly by invoking the overlap representation and Radon-transform technology, the ERBL region~\cite{Efremov:1979qk,Lepage:1980fj} can be reached.

\section{Computing the pion valence-quark GPD}
\label{sec-GPD}

A veracious description of the pion is only possible within a framework that faithfully expresses symmetries and their breaking patterns~\cite{Qin:2014vya}. The Dyson-Schwinger equations (DSEs) fulfill this requirement \cite{Chang:2011vu,Bashir:2012fs} and hence we employ that framework to compute pion properties in~\cite{Mezrag:2014jka}.
Despite its complexity, the pion bound-state is still a $J=0$ system and hence there is only one GPD associated with a quark $q$ in the pion ($\pi^\pm$, $\pi^0$), which is defined by the matrix element
\begin{equation}
\label{eq-def-GPD-H-spinless-target}
H^q_{\pi}( x, \xi, t ) =  \int \frac{\mathrm{d}^4z}{4\pi} \, e^{i x P\cdot z}\,
\delta(n\cdot z) \, \delta^2(z_\perp) \,\langle\pi(P_+)| \bar{q}\left(-z/2\right)n\cdot \gamma \; \mathcal{W}(-z/2,z/2)
q\left(z/2\right) |\pi(P_-)\rangle,
\end{equation}
where $k$, $n$ are light-like four-vectors, satisfying $k^2=0=n^2$, $k\cdot n=1$; $z_\perp$ represents that two-component part of $z$ annihilated by both $k$, $n$; and $P_\pm = P \pm \Delta/2$ 
and $\mathcal{W}(-z/2,z/2)$ represents a Wilson line laid along a light-like path that joins the two listed vectors.
In Eq.\,\eqref{eq-def-GPD-H-spinless-target}, $\xi = -n\cdot \Delta/[2 n\cdot P]$ is the ``skewness'', $t=-\Delta^2$ is the momentum transfer, and $P^2 = t/4-m_\pi^2$, $P\cdot \Delta=0$. 
The GPD also depends on the resolving scale, $\zeta$.  Within the domain upon which perturbation theory is valid, evolution to another scale $\zeta^\prime$ is described by the ERBL equations \cite{Efremov:1979qk,Lepage:1980fj} for $|x|<\xi$ and the DGLAP equations \cite{Dokshitzer:1977sg,Gribov:1972ri,Lipatov:1974qm,Altarelli:1977zs} for $|x|>\xi$, where $\xi \geq 0$.

\begin{figure}
\centering
\begin{tabular}{cc}
\includegraphics[width=6.75cm,clip]{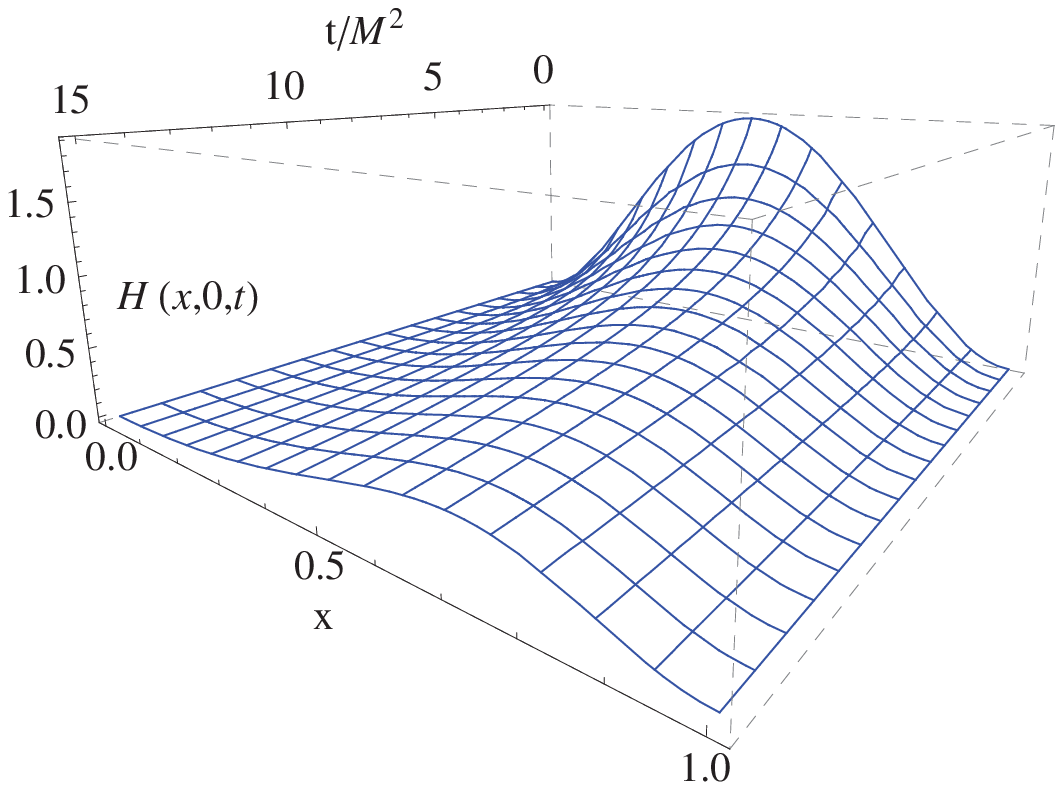} & 
\includegraphics[width=6.75cm,clip]{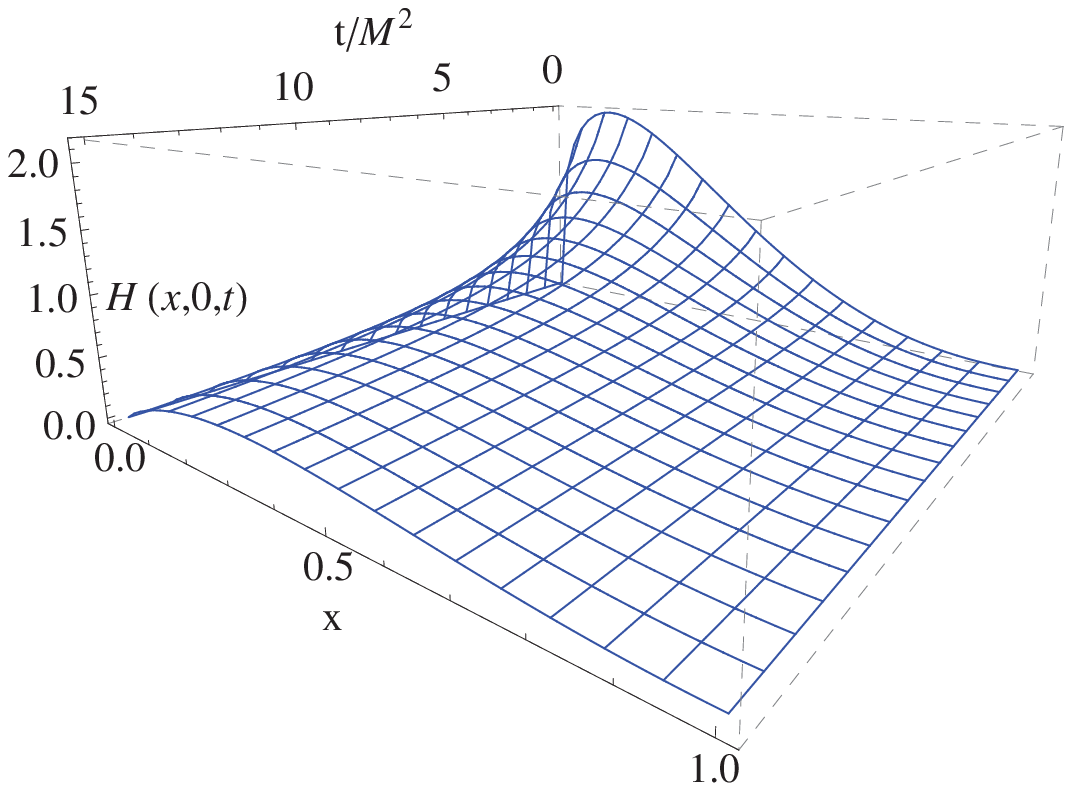}
\end{tabular}
\caption{Pion valence dressed-quark GPD, $H_\pi^{\rm v}(x,0,-\Delta_\perp^2)$, defined by the addition of Eqs.\,(\ref{eq:TriangleDiagrams},\ref{HCorrection}), obtained as explained in~\cite{Mezrag:2014jka} and plotted as a function of $t/M^2=\Delta_\perp^2/M^2$, where $M$ is a mass scale for the dressed-quark in the algebraic model of \cite{Chang:2013pq}.  \emph{Left panel}.- Result obtained at the model scale, $\zeta_H=0.51\,$GeV. \emph{Lower panel} -- GPD evolved to 
$\zeta_2=2\,$GeV using leading-order DGLAP equations.  
}
\label{fig:GPDs}       
\end{figure}

\begin{figure}
\centering
\sidecaption
\includegraphics[width=7cm,clip]{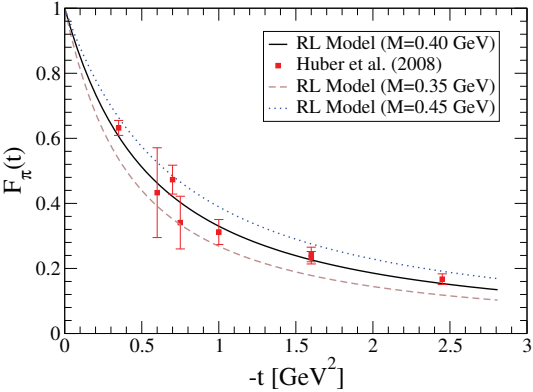}
\caption{Pion electromagnetic form factor obtained from the integtation over $x$ (sum rule) of $H_\pi^{\rm v}(x,0,-\Delta_\perp^2)$, resulting from Eqs.\,\eqref{eq:TriangleDiagrams}, \eqref{NakanishiASY} and associated definitions, in Ref.~\cite{Mezrag:2014jka}. The data are described in Ref.\,\protect\cite{Huber:2008id}. The most favourable comparison is obtained with $M=0.40\,$GeV in Eqs.\,\eqref{NakanishiASY} and the band shows results with $M=0.40\pm 0.05\,$GeV.
\label{figFpit}}
\end{figure}

As discussed at length in \cite{Mezrag:2014jka}, the valence-quark piece of the GPD expressed by \eqref{eq-def-GPD-H-spinless-target} can be first approximated by 
\begin{equation}\label{eq:TriangleDiagrams}
H_{\pi}^{\rm v}(x,\xi,t) = \frac 1 2 \;
 N_c \mathrm{tr}\rule{-0.5ex}{0ex}
 \int_{d\ell}\,\delta_n^{xP}(\ell)\,i{\Gamma}_\pi(\ell_+^{\rm R};-P_+ )\, 
S(\ell_+) \,in\cdot\Gamma(\ell_+,\ell_-) \, S(\ell_-) i\Gamma_\pi(\ell_-^{\rm R}; P_- )\, ,
\end{equation}
derived from the incomplete impulse-approximation for the so-called {\it handbag} diagram~\cite{Chang:2014lva}, and 
can be next corrected by the additional contribution
\begin{eqnarray}
H_\pi^{\rm C}(x,0,-\Delta_\perp^2) &=&
\frac{1}{2} N_c {\rm tr}\!\!\!
\int_{d\ell}\,\delta_n^{xP}(\ell)\left[
n\cdot \partial_{\ell_+^{\rm R}}\Gamma_\pi(\ell_+^{\rm R};-P_+)
S(\ell_P)\Gamma_\pi(\ell_-^{\rm R};P_-) \, S(\ell_-) \right. \nonumber \\ 
&& \left. + \Gamma_\pi(\ell_+^{\rm R};-P_+)S(\ell_P)
n\cdot \partial_{\ell_-^{\rm R}}\Gamma_\pi(\ell_-^{\rm R};P_-) S(\ell_-)
\right]\, , \label{HCorrection}
\end{eqnarray}
within the non-skewed kinematical region, for the transverse momentum $\Delta_\perp^2$. In Eqs.~(\ref{eq:TriangleDiagrams},\ref{HCorrection}), $\int_{d\ell} := \int \frac{d^4\ell}{(2\pi)^4}$ is a translationally invariant regularisation of the integral; $\delta_n^{xP}(\ell):= \delta(n\cdot \ell - x n\cdot P)$; the trace is over spinor indices; $\eta\in[0,1]$, $\bar\eta=1-\eta$; $\ell_+^{\rm R}=\bar\eta\ell_+ +\eta\ell_P$,
$\ell_-^{\rm R}=\eta\ell_- +\bar\eta\ell_P$,
$\ell_\pm = \ell \pm \Delta/2$, $\ell_P=\ell-P$ (N.B.\ Owing to Poincar\'e covariance, no observable can legitimately depend on $\eta$; i.e., the definition of the relative momentum). In order to gain novel insights into pion structure, in ref.~\cite{Mezrag:2014jka}, we used the algebraic model of \cite{Chang:2013pq}, 
\begin{subequations}
\label{NakanishiASY}
\begin{eqnarray}
\label{eq:sim1}
S(\ell) &=&[-i\gamma\cdot \ell+M]\Delta_M(\ell^2) \\
\rho_\nu(z) &=& \frac{1}{\sqrt{\pi}}\frac{\Gamma(\nu+3/2)}{\Gamma(\nu+1)}(1-z^2)^\nu \\
\label{eq:sim2}
n_\pi \Gamma_\pi(\ell^{\rm R}_\pm;\pm P) &=& i\gamma_5\int^1_{-1}dz\, \rho_\nu(z) \, \hat\Delta^\nu_M(\ell^2_{z\pm}) 
\label{NoF}
\end{eqnarray}
\end{subequations}
for the dressed-quark and pion elements in Eqs.~(\ref{eq:TriangleDiagrams},\ref{HCorrection});
where $\Delta_M(\ell^2)=1/(\ell^2+M^2)$, $M$ is a dressed-quark mass-scale; $\hat\Delta_M(\ell^2) = M^2 \Delta_M(\ell^2)$; $\ell_{z\pm}=\ell^{\rm R}_\pm + (z \pm 1) P/2$ and we work in the chiral limit ($P^2=0=\hat m$, where $\hat m$ is the current-quark mass); and $n_\pi$ is the Bethe-Salpeter amplitude's canonical normalisation constant. 
We then applied the algebra and approximations described in~\cite{Mezrag:2014tva,Mezrag:2014jka} to be finally left with the results displayed in Fig.~\ref{fig:GPDs}. Notably, the so-computed non-skewed GPD shows all the properties expected on the basis of the GPD overlap representation~\cite{Burkardt:2000za,Diehl:2000xz,Burkardt:2002hr,Diehl:2003ny}, discussed in \cite{Mezrag:2014jka},  at both the model scale, $\zeta_H=0.51\,$GeV and, after being evolved using leading-order DGLAP equations, up to  
$\zeta_2=2\,$GeV. In its forward limit, additionally, it naturally reproduces the pion valence dressed-quark distribution function found in \cite{Chang:2014lva}; and, integrated over $x$, it produces an estimate for the pion electromagnetic form factor which compares very well with experimental data (see Fig.~\ref{figFpit}). 

\section{Overlap representation, Radon transform and skewed GPD}

\begin{figure}
\centering
\sidecaption
\includegraphics[width=7cm,clip]{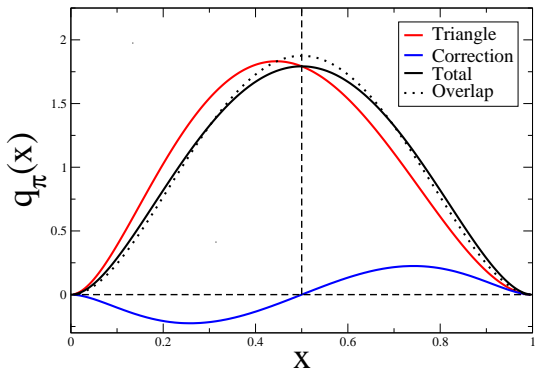}
\caption{Pion dressed-quark distribution function (PDF) obtained from the GPD in the forward limit ($t=0$, $\xi=0$); solid red and blue lines stand for the results from \eqref{eq:TriangleDiagrams} and \eqref{HCorrection}, respectively, and black solid one is for their addition; the dotted line corresponds to the overlap result given by \eqref{eq:PDF}, obtained from Eqs.~(\ref{eq:overlap},\ref{eq:LCWF}), for $\nu=1$.
}
\label{fig:PDF}       
\end{figure}

The correction \eqref{HCorrection} appears to be well defined for a non-skewed GPD. In the aim of obtaining results beyond $\xi=0$, both in DGLAP and ERBL kinematical regions, a sensible extension of \eqref{HCorrection} correcting \eqref{eq:TriangleDiagrams} is far from being obvious. A different approach based on the representation of the pion GPD as  overlap of light-cone wave functions (LCWF), 
\begin{equation}\label{eq:overlap}
H_\pi^q(x,\xi,t)_{\xi \le x \le 1} = C^q \int d^2{\bf k}_\perp^2 \Psi^\ast\left(\frac{x-\xi}{1-\xi},{\bf k}_\perp+\frac{1-x}{1-\xi}\frac{\Delta_\perp}2 ; P_- \right)
\Psi\left(\frac{x+\xi}{1+\xi},{\bf k}_\perp-\frac{1-x}{1+\xi}\frac{\Delta_\perp}2 ; P_+ \right) \ ,
\end{equation}
can be followed, where $C^q$ is a normalization constant and the LCWF can be computed by integrating over $k^-$ the pion Bethe-Salpeter wave function $\chi_\pi(k,P)$ projected onto $\gamma^+ \gamma_5$
\begin{equation}\label{eq:LCWF}
\Psi\left(k^+,{\bf k}_\perp;P\right) = - \frac 1 {2\sqrt{3}} \int \frac{dk^-}{2\pi} 
\mbox{\rm Tr}\left[ \gamma^+ \gamma_5 \, \chi_\pi(k,P)\right] \ .
\end{equation}
If we obtain the Bethe-Salpeter wave function from \eqref{NakanishiASY}, apply it to derive the LCWF with \eqref{eq:LCWF}, and use then Eq.~(\ref{eq:overlap}) to compute the GPD in the forward limit ($t=0$, $\xi=0$), we get
\begin{equation}\label{eq:PDF}
q_\pi(x) \ = \ H^q_\pi(x,0,0) \ = \ k_\nu x^{2\nu} (1-x)^{2\nu} \ .
\end{equation}
In the case $\nu=1$, the normalization factor is $k_1=30$, and the result for the pion dressed-quark distribution function (PDF) has been proved to be numerically consistent with that obtained from Eqs.~(\ref{eq:TriangleDiagrams},\ref{HCorrection}), also in the forward limit. This can be seen in Fig.~\ref{fig:PDF}, where both results appear depicted and, to the eye, can be barely distinguishable from each other. In ref.~\cite{Chang:2014lva}, the same result of \eqref{eq:PDF} with $\nu=1$ was introduced as an excellent and efficacious approximation to the pion's valence dressed-quark PDF, while also resulted, in \cite{Mezrag:2014jka}, from a heuristic LCWF implemented in \eqref{eq:overlap}. Here, it appears as the natural result from the algebraic model of Eqs.~\eqref{NakanishiASY}~\cite{Chang:2013pq} and the overlap representation of pion's GPD. 

Therefore, Eq.~(\ref{eq:overlap}) provides with a natural extension for the GPD beyond the non-skewed limit, although only valid within the DGLAP region. Moreover, one can take also advantage of that, within the one-component DD (1CDD) scheme (see discussion in~\cite{Mezrag:2013mya} and references therein), the GPD can be expressed as the 
Radon transform, $\mathcal{R}f$, of the distribution $f(\beta,\alpha,t)$,
\begin{equation}
\label{eq:radon}
\frac{\sqrt{1+\xi^2}}{x} \, H(x,\xi,t) = \mathcal{R}f(s,\varphi,t) = \int_{|\alpha|+|\beta|\leq 1} \rule{-5ex}{0ex} d\beta \,  d\alpha \,
\delta(s-\beta \cos{\varphi}-\alpha\sin{\varphi})\, f(\beta,\alpha,t) \,,
\end{equation}
where $s=x\cos{\varphi}$ and $\xi=\tan{\varphi}$. Then, the singular value decomposition of the Radon transform defined in \eqref{eq:radon}, investigated within the context of the computerized tomography~\cite{natterer:2001}, allows for the GPD to be recast as
\begin{equation}\label{eq:fin}
\mathcal{R}f(s,\varphi,t) \ = \ \sum_{m=0}^\infty \sum_{l=0}^m g_{ml}(t) \ e^{i\; (-m+2l) \; \varphi} \, C^\alpha_m(s) \ ,
\end{equation}
where $g_{ml}(t)$ are the coefficients for the expansion in terms of Geigenbauer polynomials, $C^\alpha_m(s)$. Therefore, identifying properly the coefficients $g_{ml}(t)$, one is also left with the natural extension to the ERBL kinematical region for the GPD, {\it via} \eqref{eq:fin}. This might pave the way to build a model implementing, {\it a priori}, positivity and polynomiality properties. 

\section{Conclusions}

We have briefly reported on the first steps recently made towards the computation of the pion's valence dressed-quark GPD, within the symmetry-preserving framework provided by DSEs. In addition, and very preliminary, we also discussed the open avenue, on the basis of the GPD representation as overlap of light-cone wave functions, and that can be exploited to extend previous results; in particular, from those for the GPD in DGLAP kinematical region and near the forward limit to non-zero skewness and to the ERBL domain.

\begin{acknowledgement}

This contribution is based on research performed in an ongoing collaboration with L. Chang and C.D.~Roberts and the work has been partially supported by Spanish ministry projects FPA-2011-23781 and FPA-2014-53631-C2-2-P, French GDR 3034 PH-QCD ``Cromodynamique Quantique et Physique des Hadrons" and ANR-12-MONU-0008-01 ``PARTONS" and Argonne National Laboratory, LDRD project 2016-188-N0 ``Understanding the Structure of Matter'',  and U.S. Department of Energy, Office of Science, Office of Nuclear Physics, under contract no.~DE-AC02-06CH11357.  We also thank for their very valuable remarks to F.~Sabati\'e and S.M.~Schmidt. H. M. and J. R-Q. are grateful for the chance to participate in the ``21st Conference on Few-Body Problems in Physics", held in Chicago, and specially to C.D.~Roberts for his warm hospitality at the Argonne National Laboratory. 

\end{acknowledgement}

%
\bibliography{RodriguezQuinteroJ.bib}
%
%
%
%

\end{document}